\documentstyle[12pt]{article}
\def \F{\phi}

\def\NP{{\it Nucl. Phys.\ }}

\def\PL{{\it Phys. Lett.\ }}
\def\PR{{\it Phys. Rev.\ }}

\def\r{\rho}

\def\b{\beta}
\def\a{\alpha}

\def\s{\sigma}

\def\half{{1\over 2}}
\def\d{\dagger}
\def\be{\begin{equation}}
\def\eq{\end{equation}}
\def\Tr{{\rm Tr}}

\def\q2{{\rm QCD}_{2}}
\def\q4{{\rm QCD}_{4}}

\begin{document}

\begin{flushright}
OUTP-9548P\\
DAMTP-95-71\\
hep-th/9512106\\
\end{flushright}
\vspace{20mm}
\begin{center}
{\LARGE  A $(1+1)$-Dimensional Reduced Model of Mesons\\}
\vspace{30mm}
${\rm {\bf F. Antonuccio}}^1$  and ${\rm {\bf S. Dalley}}^2$\\
\vspace{5mm}
${}^1${\em Theoretical Physics\\
1 Keble Road, Oxford, U.K.}\\
\vspace{5mm}
${}^2${\em Department of Applied Maths and Theoretical Physics\\
Silver Street, Cambridge, U.K.}\\
\end{center}
\vspace{30mm}

\abstract
We propose an extension of 't Hooft's large-$N_c$  light-front
QCD in two dimensions to include helicity and physical gluon degrees
of freedom, modelled on a classical
dimensional reduction of  four dimensional QCD. A
non-perturbative
renormalisation of the infinite set of coupled integral equations
describing boundstates is performed.
These equations are then solved, both analytically
in a phase space wavefunction approximation and numerically by
discretising
momenta, for (hybrid) meson masses and (polarized) parton structure functions.

\newpage
\baselineskip .25in
\section{Introduction}

't Hooft suggested a two-dimensional model of mesons \cite{hoof} which
is a tractable non-abelian gauge theory sharing many qualitative
properties with true QCD in four dimensions. The keys to its solution
involved the use of the large $N_c$ limit and light-front quantisation in
light-front gauge. Since it is two dimensional, it contains no spin or
physical gluons however; these are perhaps the least well understood
structure of true hadrons. One may retain a remnant of such transverse
features,
whilst keeping the kinematics $1+1$-dimensional, by picking two
(arbitrary)
space dimensions $x_{\perp}= \{ x^1 , x^2 \}$
and considering only the zero modes
\be
\partial_{x_{\perp}} A_{\mu} = \partial_{x_{\perp}} \Psi  = 0 \label{tzero}
\eq
of the gauge and quark fields.
This may well be an appropriate approximation in some high energy
scattering
processes when transverse momenta are relatively damped. With a view
to
employing this fact, in this paper we construct the mesons appearing
in a $1+1$-dimensional reduced model specified by imposing
(\ref{tzero})
at the classical level,
which will appear as asymptotic or exchanged states (in the Regge
sense) for scattering processes in this model.\footnote{We have
performed a detailed analysis
of the corresponding pure glue states in a recent paper
\cite{us}, building upon earlier work \cite{dk}.
Similar work, with a truncated light-front Fock space,
has recently appeared for $N_c =3$ \cite{su3}} These states need not be similar
to the true hadrons involved in high energy scattering,
since
solving the  boundstate problem for only co-linear
($k_{\perp} = 0$) quarks and gluons is not the same as solving the
full
boundstate problem and then going to a regime where $k_{\perp}$ is
relatively
small.
However, the results presented here and in ref.\cite{us} encourage
us that the hadrons reduced in this way
share many qualitative properties with true hadrons,
even those Lorentz-invariant ones for which $k_{\perp} =0$ cannot be
kinematically
justified, and
therefore are of interest in their own right.

In the large-$N_c$ light-front
formalism the reduced hadrons satisfy infinite sets of
coupled boundstate integral equations
which, we argue, are rendered finite equation-by-equation by retaining
parton
self-energies. Solutions to the equations are obtained
analytically
in a phase space wavefunction approximation and numerically by
truncating and
discretising them.
These solutions yield both meson and hybrid meson masses, as well as
(polarized) quark and gluon structure functions.

\section{Dimensional Reduction.}

We start from $SU(N_c)$ gauge theory in $3+1$-dimensions ($\mu \in
\{ 0,1,2,3\} $) with one flavour of quarks
\be
S = \int d^4 x  \left[ -{1\over 4 \tilde{g}^2} \Tr F_{\mu\nu} F^{\mu\nu}
+ {\rm i}\bar{\Psi} \gamma^{\mu}_{(4)} D_{\mu} \Psi - m \bar{\Psi}
\Psi \right]
\eq
in the Weyl representation
\begin{eqnarray}
\gamma^{0}_{(4)} = \left( \begin{array}{cc}
 0 &  -\bf{1} \\ -\bf{1} & 0  \end{array} \right) \ , \
 \gamma^{i}_{(4)} = \left( \begin{array}{cc}
0 & \s^{i} \\ -\s^i & 0 \end{array} \right) \ .
\end{eqnarray}
Imposing (\ref{tzero}),
one finds an
effectively two-dimensional gauge theory of adjoint scalars and
fundamental Dirac spinors with action
\begin{eqnarray}
S & = & \int dx^0 dx^3 \ \left\{ -{1 \over 4g^2} \Tr[F_{a b}F^{ab}]+
{{\rm i} \over \sqrt{2}} (\bar{u} \gamma^{a}_{(2)}
D_{a} u + \bar{v} \gamma^{a}_{(2)} D_{a} v) + {m\over \sqrt{2}} (\bar{u} v
+ \bar{v} u) \right. \nonumber  \\
& & + \Tr\left[ -\half \bar{D}_{a}
\F_{\r} \bar{D}^{a} \F^{\r} -{tg^2 \over 4}
[\F_{\r},\F_{\s}][\F^{\r},\F^{\s}]
+ \half m_{0}^{2} \F_{\r}\F^{\r} \right]
\nonumber  \\
 & & \left. - {sg\over \sqrt{2}} (\bar{u} (\F_1 + {\rm
i}\gamma^{5}_{(2)} \F_2)
u - \bar{v}(\F_1 - {\rm i} \gamma^{5}_{(2)} \F_2 )v) \right\} \ ,\label{red}
\end{eqnarray}
where $a$ and $b \in \{ 0,3\}$, $\r \in \{ 1,2 \}$,
$\gamma^{0}_{(2)} = \s^1$, $\gamma^{3}_{(2)} = {\rm i} \s^2$,
$\gamma^{5}_{(2)} = {\rm i} \s^1 \s^2$,
$\F_{\r} = A_{\r}/g$, $g^2 = \tilde{g}^2 / \int dx^1 dx^2$,
$\bar{D}_{a} = \partial_{a} + {\rm i}[A_{a},.]$, $D_{a} =
\partial_{a} + {\rm i}A_{a}$.
The two-component
spinors $u$ and $v$ are related to $\Psi$ by
\begin{eqnarray}
2^{1/4} \Psi \sqrt{\int dx^1 dx^2} = \left( \begin{array}{c}
 u_{R+} \\  u_{L+} \\  u_{L-} \\  u_{R-}  \end{array} \right)
\ , \
 u = \left( \begin{array}{c}
u_{L+} \\ u_{R+} \end{array} \right) \ , \ v = \left( \begin{array}{c}
u_{L-} \\ u_{R-} \end{array} \right) \ .
\label{spin}
\end{eqnarray}
The suffices $L$ $(R)$ and $+$ $(-)$ in (\ref{spin})
represent Left (Right) movers and
$+ve$ $(-ve)$ helicity.
Thus $u,v,\F_1 ,\F_2$ represent the transverse polarisations of the
$3+1$
dimensional quarks and gluons. Since the dimensional reduction
procedure
treats space asymmetrically, we have allowed the  gauge
couplings in transverse directions to be different from that in the
longitudinal direction in general, through dimensionless parameters
$s$ and $t$ (canonically $s=-t =1$), and have added a bare mass
$m_0$ for
the $\F$ fields; both these modifications can occur due to loss of
transverse local gauge transformations.
The fact that the reduction to colinearity has been performed at the
classical level, avoiding the singularities of the corresponding
procedure performed after quantisation, emphasizes
that this is a {\em model}.
The couplings and masses are therefore left as adjustable parameters;
for example, the $s$ and $t$ couplings will control
the
magnitude of helicity splittings in the spectrum.

The action (\ref{red}) is a combination of gauged fundamental
\cite{hoof} and
adjoint matter \cite{dk} representations with further Yukawa \cite{brod,perry}
and matrix scalar \cite{dk2} interactions,
each of which have been individually
studied by light-front quantisation before. We therefore only
sketch
the construction of boundstate equations. These equations will be
valid for
the  modes of the theory at $N_c =\infty$
with non-zero light-front momenta $k^+ = (k^0
+ k^3)/\sqrt{2} \neq 0$. Zero $k^+$ modes in reduced models
have been discussed in refs.\cite{su3,zero} for example, and while
there is no  evidence that they affect the spectrum in the
two-dimensional large-$N_c$ theory, one should bear
in mind their omission in this initial investigation.
In the naive
light-front  gauge $A_- = (A_0 - A_3)/\sqrt{2} = 0$, the fields
$A_+$ and $u_{L \pm}$ are non-propagating in light-front time $x^+ =
(x^0 + x^3)/\sqrt{2}$, satisfying constraint equations
\begin{eqnarray}
0 & = & {\rm i} \partial_{-} u_{L+} + {m \over \sqrt{2}}u_{R-} - sg
 B_+
u_{R+} \\
0 & = & {\rm i} \partial_{-} u_{L-} + {m \over \sqrt{2}}u_{R+} +
sg B_-
u_{R-} \\
0 & = & \partial_{-}^{2} A_{+} -  g^2 J^+   \label{six}
\end{eqnarray}
where the helicity fields and longitudinal momentum current are
\begin{eqnarray}
B_{\pm} & = & (\F_1 \pm {\rm i} \F_2)/\sqrt{2} \\
J^+_{ij} & = &  {\rm i} [B_-, \partial_{-} B_+ ]_{ij} + {\rm i} [B_+,
\partial_{-} B_- ]_{ij}  + u_{R+i}u^{*}_{R+j} + u_{R-i}u^{*}_{R-j} \ .
\label{eight}
\end{eqnarray}
In eq(\ref{eight}) we have explicitly
shown the colour indices $i,j = 1,\ldots,N_c$.
Eliminating the constrained fields in favour of non-local
instantaneous
interactions gives the light-front energy and momentum
\begin{eqnarray}
P^- & = &
    \int dx^- \left[{tg^2\over 2} (B_{+ij}B_{+jk}B_{-kl}B_{-li}-
B_{+ij}B_{-jk}B_{+kl}B_{-li})
- {g^2 \over 2} J^{+}_{ij} {1 \over \partial_{-}^{2}} J^{+}_{ji}
+  \half m_{0}^{2} B_{+ij}B_{-ji} \right. \nonumber  \\
& & - {{\rm i} m^2 \over 2} \left( u^{*}_{-i} {1 \over \partial_{-}}
u_{-i} + u^{*}_{+i} {1 \over \partial_{-}} u_{+i} \right)
-{\rm i} s^2 g^2  \left( u^{*}_{+i} B_{-ij} {1 \over \partial_{-}}
B_{+jk} u_{+k} + u^{*}_{-i} B_{+ij} {1 \over \partial_{-}}
B_{-jk} u_{-k}\right) \nonumber \\
&& \left.
 -{{\rm i} sgm \over \sqrt{2}} ( u^{*}_{+i} [\partial^{-1}_{-},
B_{-ij}] u_{-j} + u^{*}_{-i} [B_{+ij}, \partial^{-1}_{-}] u_{+j})
 \frac{}{} \right]
\label{minus}\\
P^+  & = & \int dx^-
   \left[ 2 \partial_{-} B_{+ij} \partial_{-} B_{-ji}
+ {\rm i} ( u^{*}_{+i} \partial_{-} u_{+i} + u^{*}_{-i} \partial_{-}
u_{-i}) \right]
\label{plus}
\end{eqnarray}
where the right-mover subscript $R$ on $u$ has been dropped.
The fourier modes defined at $x^+ =0$ by\footnote{symbol $\d$ is the
quantum version of complex conjugate $*$ and does not act on colour indices.}
\begin{eqnarray}
u_{\pm i} & =& {1 \over \sqrt{2\pi}} \int_{0}^{\infty} dk^+
\left( b_{\pm i}(k^+) {\rm e}^{-ik^+x^-}
+ d^{\d}_{\mp i}(k^+) {\rm
e}^{ik^+x^-} \right)  \label{mode1}\\
B_{\pm ij} & =& {1 \over \sqrt{2\pi}} \int_{0}^{\infty} {dk^+ \over
\sqrt{2k^+}} \left( a_{\mp ij}(k^+) {\rm e}^{-ik^+x^-} + a_{\pm
ji}^{\d}(k^+) {\rm e}^{ik^+x^-} \right) \label{mode2}
\end{eqnarray}
satisfy
\be
\{ b_{\a i}(k^+),b_{\b j}^{\d}(\tilde{k}^+) \} = \delta (k^+ -
\tilde{k}^+) \delta_{ij} \delta_{\a \b} \ , \
[a_{\a ij}(k^+), a_{\b kl}^{\d}(\tilde{k}^+)] =\delta (k^+ -
\tilde{k}^+) \delta_{ik} \delta_{jl} \delta_{\a \b}
\label{osc}
\eq
when the canonical equal-$x^+$ relations hold
\be
\{ u^{*}_{\a i}(x^-), u_{\b j}(\tilde{x}^-) \} =
\delta( x^- -
\tilde{x}^- )
\delta_{ij} \delta_{\a \b} \ ,  \ [\F_{\a ij} (x^-), \partial_{-}
\F_{\b kl}
(\tilde{x}^-) ] = {{\rm i} \over 2} \delta( x^- -
\tilde{x}^- ) \delta_{il} \delta_{jk} \delta_{\a \b} \ ,
\eq
where $\a$ and  $\b \in \{+,-\}$ label helicity.
The light-front Fock vacuum is well known to be trivial, satisfying
$:P^+ : |0> = :P^- : |0> =0$,  so Fock states built with the
oscillators (\ref{osc}) have a direct partonic interpretation.
In the large-$N_c$ limit with $g^2 N_c$ fixed,
the finite solutions to the mass-shell condition $M^2 \Psi = 2 P^+
P^- \Psi$  are formed from singlet
linear
combinations of the Fock states under residual $x^-$-independent gauge
transformations
\begin{eqnarray}
\Psi & = & \sum_{n=2}^{\infty} \int_{0}^{P^+} dk_1^+ dk_2^+ \dots
dk_n^+ \hspace{1mm} \delta \left(k_1^+ + k_2^+ + \cdots + k_n^+ - P^+
\right) \times \nonumber \\
& & \frac{f_{\alpha \beta \gamma \dots \eta
\delta}(k_1^+,\dots,k_n^+)}{\sqrt{N_{c}^{n-1}}}d^{\d}_{\alpha i }(k_1^+)
      a^{\d}_{\beta i j }(k_2^+) a^{\d}_{\gamma j k}(k_3^+)
        \dots a^{\d}_{\eta l p}(k_{n-1}^+) b^{\d}_{\delta p}(k_n^+)
               |0>,  \label{singletstates}
\end{eqnarray}
Only these singlets satisfy the
quantum version of the zero mode  of eq(\ref{six}),
$\int dx^-  :J^+: \Psi = 0$. They are also
eigenstates of the total helicity
\be
h= \int_{0}^{P^+}  dk^+ \   \sum_{\a}
{\rm sgn}(\a)\left[ \half (b^{\d}_{\a i}
b_{\a i}(k^+)
+ d^{\d}_{\a i} d_{\a i} (k^+)) + a^{\d}_{\a ij} a_{\a ij} (k^+)
\right] \ ,
\eq
forming degenerate pairs $\pm h$ when $h \neq 0$, and total momentum
$P^+$.
It remains therefore to find the coefficients $f$ which diagonalize $P^-$.

\section{Boundstate Equations.}
Substituting the mode expansions (\ref{mode1})(\ref{mode2}) into
$P^{\pm}$ (\ref{minus})(\ref{plus})
and  discarding infinite additive constants but {\em not} quadratic
terms resulting from normal ordering one finds
\be
P^+ = \int_{0}^{\infty} dk \ k   \left(b^{\d}_{\a i}
b_{\a i}(k)
+ d^{\d}_{\a i} d_{\a i} (k) + a^{\d}_{\a ij} a_{\a ij} (k) \right)
\eq
and $P^- = P^-_{{\rm trans.}} + P^-_{{\rm long.}} + P^-_{{\rm glue}}$,
where
\begin{eqnarray}
 P^-_{{\rm trans.}} & = & \frac{s^2 g^2}{2 \pi } \int_{0}^{\infty}
                                dk_1 dk_2 dk_3 dk_4
       \left\{ \frac{\delta(k_1 + k_2 - k_3 - k_4)}{\sqrt{k_1 k_3}
        (k_1+k_2)} \times  \right. \nonumber
                       \\
& & \left( a^{\d}_{+ik}(k_3)b^{\d}_{-k}(k_4)
                          a_{+il}(k_1)b_{-l}(k_2) +
                           d^{\d}_{+k}(k_4)a^{\d}_{-kj}(k_3)
                           d_{+l}(k_2)a_{-lj}(k_1) \right)  \nonumber \\
&  & +  \frac{\delta(k_1 + k_2 + k_3 - k_4)}{\sqrt{k_1 k_2}
        (k_2+k_3)} \times \nonumber
                   \\ & & \left( b^{\d}_{+i}(k_4)a_{+ik}(k_1)
                          a_{-kl}(k_2)b_{+l}(k_3) +
                           d^{\d}_{+i}(k_4)d_{+l}(k_3)
                      a_{-lk}(k_2)a_{+ki}(k_1)  \right. \nonumber \\
& &  \left.  + a^{\d}_{+ik}(k_1)a^{\d}_{-kl}(k_2)b^{\d}_{+l}(k_3)
             b_{+l}(k_4) +
     d^{\d}_{+i}(k_3)a^{\d}_{-ik}(k_2)a^{\d}_{+kl}(k_1)d_{+l}(k_4)
        \right)  \nonumber \\
&  & +  \left. \hspace{4mm} ({\rm interchange}
       \hspace{2mm} + \leftrightarrow - ) \hspace{8mm}
                 \frac{\hbox{}}{\hbox{}} \! \! \! \right\}  \nonumber \\
& + &  \frac{m s g}{2\sqrt{ \pi}} \int_{0}^{\infty} dk_1 dk_2 dk_3
     \left\{ \frac{}{} \delta(k_1 + k_2 - k_3) \frac{1}{\sqrt{k_1}} \left(
        \frac{1}{k_2} - \frac{1}{k_3} \right) \times
           \right. \nonumber \\
& & \left( b^{\d}_{+i}(k_3)a_{+ij}(k_1)b_{-j}(k_2) -
           d^{\d}_{+i}(k_3)d_{-j}(k_2)a_{+ji}(k_1) \right.
                  \nonumber \\
& & + \left. a^{\d}_{+ij}(k_1)b^{\d}_{-j}(k_2)b_{+i}(k_3) -
          d^{\d}_{-i}(k_2)a^{\d}_{+ij}(k_1)d_{+j}(k_3) \right)
                   \nonumber \\
& & -  \left. \hspace{4mm} ({\rm interchange}
       \hspace{2mm} + \leftrightarrow - ) \hspace{8mm}
               \frac{\hbox{}}{\hbox{}} \! \! \! \right\}
                   \nonumber \\
&+& \half m^2 \int_{0}^{\infty} \frac{dk}{k}
       \left(
             d^{\d}_{\a i}(k)d_{\a i}(k) + b^{\d}_{\a i}(k)b_{\a i}(k) \right)
      \nonumber \\
& + &  \frac{s^2 g^2 N_c}{ 2 \pi} \int_{0}^{\infty}
  \frac{dk_1 dk_2}{k_1 (k_2-k_1)} \left(
     b^{\d}_{\a i}(k_2)b_{\a i}(k_2)  \right) \nonumber \\
& + & \frac{s^2 g^2 N_c}{ 2 \pi} \int_{0}^{\infty}
  \frac{dk_1 dk_2}{k_1 (k_2+k_1)} \left(
     d^{\d}_{\a i}(k_2)d_{\a i}(k_2)  \right) \label{ptrans} ,
\end{eqnarray}

\begin{eqnarray}
 P^-_{{\rm long.}} & = & \frac{g^2}{2\pi}
  \int_{0}^{\infty} dk_1 dk_2 dk_3 dk_4
                        \hspace{1mm} \left\{
    \frac{  \delta (k_1+k_2-k_3-k_4)}{ (k_2 - k_4)^2 } \hspace{1mm}
         d^{\d}_{\alpha i}(k_3)b^{\d}_{\beta j}(k_4)
           d_{\alpha i}(k_1)b_{\beta j}(k_2)
   \right.  \nonumber \\
& - & \delta (k_1+k_2-k_3-k_4) \hspace{1mm}
\frac{ (k_1+k_3)}{ 2 \sqrt{k_1 k_3} (k_2-k_4)^2} \times
     \nonumber \\
& &  \left(  a^{\d}_{\alpha ij}(k_3)b^{\d}_{\beta j}(k_4)
           a_{\alpha ik}(k_1)b_{\beta k}(k_2) +
           d^{\d}_{\alpha j}(k_4)a^{\d}_{\beta ji}(k_3)
           d_{\alpha k}(k_2)a_{\beta ki}(k_1) \right)
      \nonumber  \\
& + & \delta (k_1+k_2+k_3-k_4) \hspace{1mm}
        \frac{k_1 - k_2}{2 \sqrt{k_1 k_2} (k_1+k_2)^2} (1-\delta_{\a
\gamma} ) \times
       \nonumber \\
& & \left( a^{\d}_{\alpha ij}(k_1) a^{\d}_{\gamma jk}(k_2)
             b^{\d}_{\beta k}(k_3) b_{\beta i}(k_4) +
           d^{\d}_{\beta k}(k_3) a^{\d}_{\alpha kj}(k_2)
           a^{\d}_{\gamma ji}(k_1) d_{\beta i}(k_4) \right.
    \nonumber \\
& & + \left. \left. b^{\d}_{\beta i}(k_4)a_{\alpha ij}(k_1)
             a_{\gamma jk}(k_2) b_{\beta k}(k_3) +
      d^{\d}_{\beta i}(k_4) d_{\beta k}(k_3)
           a_{\alpha kj}(k_2) a_{\gamma ji}(k_1)  \right)
      \frac{\hbox{}}{\hbox{}} \! \! \!  \right\} \nonumber \\
&+& \frac{g^2 N_c}{4 \pi} \int_{0}^{\infty} dk_1 dk_2
 \left( \frac{1}{(k_1-k_2)^2} - \frac{1}{(k_1+k_2)^2} \right)
      \times \nonumber \\
& &  \hspace{35mm} \left( d^{\d}_{\beta i}(k_1)d_{\beta i}(k_2) +
                 b^{\d}_{\beta i}(k_1)b_{\beta i}(k_2) \right),
      \label{plong}
\end{eqnarray}
and
\begin{eqnarray}
 P^-_{{\rm glue}}  & = &
 \frac{g^2}{2 \pi} \int_{0}^{\infty} \frac{dk_1 dk_2 dk_3 dk_4}
          {\sqrt{k_1 k_2 k_3 k_4}} \hspace{1mm}
  \left\{ \frac{}{} \delta(k_1 + k_2 -k_3 -k_4)
           \times    \right. \nonumber \\
& &  \left[ \frac{}{}
 (t-A_2) \hspace{1mm} a^{\d}_{+ ij}(k_3) a^{\d}_{+ jk}(k_4)
                a_{+ il}(k_1) a_{+ lk}(k_2)   \right.
     \nonumber \\
& & + (t+A_1) \hspace{1mm}  a^{\d}_{+ ij}(k_3) a^{\d}_{- jk}(k_4)
                a_{- il}(k_1) a_{+ lk}(k_2) \nonumber \\
& & + (A_1 + A_2 - 2t)  \hspace{1mm}
          a^{\d}_{+ ij}(k_3) a^{\d}_{- jk}(k_4)
                a_{+ il}(k_1) a_{- lk}(k_2)
      \left.           \frac{}{} \right] \nonumber \\
&+&  \delta(k_1 + k_2 + k_3 -k_4)  \left[ \frac{}{}
         \right. (B_1+B_2-2t) \times \nonumber \\
& &  \hspace{1mm} \left(
       a^{\d}_{+ ij}(k_1) a^{\d}_{- jk}(k_2)
      a^{\d}_{+ kl}(k_3) a_{+ il}(k_4) +
      a^{\d}_{+ ij}(k_4) a_{+ il}(k_1) a_{- lk}(k_2)
                                  a_{+ kj}(k_3)
 \nonumber \right) \nonumber \\
& & + (t+B_1) \hspace{1mm} \left(
     a^{\d}_{+ ij}(k_1) a^{\d}_{- jk}(k_2)
      a^{\d}_{- kl}(k_3) a_{- il}(k_4) +
      a^{\d}_{- ij}(k_4) a_{+ il}(k_1) a_{- lk}(k_2)
                                  a_{- kj}(k_3) \right)
           \nonumber \\
& & +(t+B_2) \hspace{1mm} \left(
    a^{\d}_{+ ij}(k_1) a^{\d}_{+ jk}(k_2)
      a^{\d}_{- kl}(k_3) a_{+ il}(k_4) +
      a^{\d}_{+ ij}(k_4) a_{+ il}(k_1) a_{+ lk}(k_2)
                                  a_{- kj}(k_3) \right)
 \left. \frac{}{} \right] \nonumber
\\
& + &
     \left[ \hspace{4mm} ({\rm interchange}
       \hspace{2mm} + \leftrightarrow - ) \hspace{8mm}
          \right] \hspace{1mm}
       \left. \frac{}{} \right\}   \hspace{3mm}
 +  \hspace{3mm} \frac{g^2 N_c}{\pi} \int_{0}^{\infty} {dk_2 \over
k_2} \int_{0}^{k_2} dk_1  \frac{1}{(k_2-k_1)^2} a^{\d}_{\alpha i j}(k_2)
                          a_{\alpha i j}(k_2) \nonumber \\
& + & \hspace{3mm}  \frac{g^2 N_c(1-t/2)}{4\pi}
\int_{0}^{\infty} {dk_1 dk_2 \over k_1 k_2} a^{\d}_{\alpha i j}(k_2)
                          a_{\alpha i j}(k_2)
,  \label{pglue}
\end{eqnarray}
where the coefficients $A_1,A_2,B_1,B_2$ are defined by
\begin{eqnarray}
 A_1 = \frac{(k_2-k_1)(k_4-k_3)}{4 (k_1+k_2)^2}, & &
 A_2 = \frac{(k_1+k_3)(k_2+k_4)}{4 (k_4-k_2)^2}, \\
 B_1 = \frac{(k_1-k_2)(k_3+k_4)}{4 (k_1+k_2)^2}, & &
 B_2 = \frac{(k_1+k_4)(k_3-k_2)}{4 (k_2+k_3)^2}.
\end{eqnarray}
A simplification of large $N_c$ is that $2 \leftrightarrow 2$ or $1
\leftrightarrow 3$ interactions occur only between
{\em neighboring} partons in the colour contraction
(\ref{singletstates}). Pair
production of quarks, but not gluons, is  absent.
The last integral in (\ref{plong}), last two
integrals in (\ref{ptrans}), and last two integrals in (\ref{pglue}) are
$x^+$-instantaneous
self-energy terms (refered to as $s$,$t$, and $g$ self-energy
hereafter)
resulting from normal ordering, which lead to a finite spectrum if
retained. This is most easily demonstrated by
projecting
$M^2 \Psi$ onto individual Fock states to derive
integral equations for the coefficients $f$. One obtains infinite
towers
of successively coupled
equations that are too long to write in full detail here, so we exhibit
salient properties only.

\subsection{Longitudinal.}
In this section we turn off the transverse
couplings $s=t=0$, so that only the number $n$ of
partons is significant rather than their individual helicities. If we
also neglect the longitudinal
processes which change the number of partons, which
is in fact an excellant approximation for low eigenvalues $M$ \cite{dk},
then for $f_n(x_1,\ldots,x_n)$, where $x_i = k_{i}^{+}/P^{+}$ labels
momentum fraction, one finds
\begin{eqnarray}
{M^2 \pi \over g^2 N_c} f_n  & = & {m_{B}^{2}\pi \over
g^2 N} [{1 \over x_2} + \cdots + {1 \over x_{n-1}}] f_n  +
{m^{2} \pi \over g^2 N} [{1 \over x_1}+ {1 \over x_n}] f_n
+ \delta_{n2} \int_{0}^{1} dy
\left\{ {f_2(x_1,x_2) - f_2(y,1-y) \over (y-x_1)^2}	\right\}
\nonumber \\
&& \nonumber \\
&+&  \sum_{i=2}^{n-2} \int_{0}^{x_i + x_{i+1}} dy {(x_i + y)(x_i +
2x_{i+1} -y) \over 4(x_i -y)^2 \sqrt{x_i x_{i+1} y (x_i + x_{i+1}-y)}}
\{  f_n(x_1, \ldots ,x_n)   \nonumber \\
&&  - f_n(x_1,\ldots,x_{i-1},y,x_i + x_{i+1} -y, \ldots,x_n) \} +
{\pi \over 4 \sqrt{x_i x_{i+1}}} f_n \label{bs} \\
&&\nonumber \\
&+& \int_{0}^{x_{n-1} + x_{n}} dy  {x_{n} +2x_{n-1} -y
\over 2(x_n -y)^2\sqrt{x_{n-1}(x_n +x_{n-1} -y)}} \{
f_n (x_1,\ldots , x_n)
\nonumber \\
&&  - f_n(x_1,\ldots,x_n+x_{n-1}-y,y) \}  +  {1\over
x_n}
\left(  \sqrt{1 + {x_n \over x_{n-1}}} -1 \right) f_n  \nonumber \\
&&\nonumber \\
&+& \int_{0}^{x_1 + x_2} dy {x_1 +2x_2 -y \over 2(x_1 -y)^2\sqrt{x_2
(x_1 +x_2 -y)}} \{ f_n(x_1 , \dots, x_n)   \nonumber \\
&&  - f_n(y,x_1+x_2-y,\ldots,x_n) \}
+ {1\over x_1 }\left( \sqrt{1 + {x_1\over x_2} } -1 \right) f_n
\nonumber \ .
\end{eqnarray}
The longitudinally induced $g$ self-energy  for fermions and gluons has a
linear divergent piece which has been cancelled against the poles in $y$ in
each of the above integrals (curly brackets).
These represent the $x^+$-instantaneous Coulomb exchange
of  $A_+$ quanta between  partons,
which diverges for zero exchanged momentum.
The self-energy of the gluon also contains a logarithmic divergence
which may be
absorbed by the bare mass $m_0$ to leave renormalised mass
$m_B$.
The way we have written the two-body Coulomb kernels
above means that certain finite parts are left over after cancellation
of divergences and renormalisation of masses, i.e. we identify the following
static interactions between two partons depending upon the statistics and
momentum
fraction of each
\begin{eqnarray}
&{\rm quark}(x_1), {\rm gluon}(x_2)& \ \ \ {1 \over x_1}\left(\sqrt{1
+ {x_1 \over x_2}} -1 \right) \label{qg}\\
&{\rm gluon}(x_i),{\rm gluon}(x_{i+1})& \ \ \ {\pi \over 4\sqrt{x_i
x_{i+1}}}\label{gg} \\
&{\rm quark},{\rm anti-quark}& \ \ \ 0 \ . \label{qq}
\end{eqnarray}
Since these are interactions between partons with contracted colour
indices,
it is useful to interpret them as groundstate energy of the flux
line between partons. The Coulomb integrals describe the
longitudinal excitations of the flux line in each case. For example,
with
no gluons ($n=2$) one recovers t'Hooft's  model \cite{hoof}, the zero
groundstate energy (\ref{qq}) resulting in a massless meson when
$m \to 0$.
When $m_B \to m \to 0$, a good approximation to
the groundstate for each $n$ is given by the
phase space wavefunction $\psi_n$, defined by the ansatz $f_{n'} =
{\rm const.}\delta_{nn'}$.
The integrals shown and in fact all other longitudinal processes
neglected in (\ref{bs}) vanish for this ansatz. For $n-2>0$ gluons one
finds that $M^2$ is diagonal in the sub-basis of $\psi_n$'s
\be
M^2 |\psi_n> = g^2 N \left( {((n-2)^2-1) \pi\over 4} + {4(n-1)
\log{2}\over
\pi}\right) |\psi_n> \ ,
\eq
where the first term comes from  gluon-gluon (\ref{gg}) and the
second term from quark-gluon (\ref{qg}) flux energy. For small $n$ we
found
that this formula  typically differs from a numerical
solution by about $10 \%$.
Each $\psi_n$ is $2^n$-fold degenerate in helicities and forms a
groundstate on top of which there exists a discrete trajectory of
Coulomb excitations just as for $n=2$ \cite{hoof}.
For high enough $M$ however, this picture appears to break down
since the processes which mix sectors of different $n$, neglected in
(\ref{bs}), become important \cite{dk}. But we emphasize  that the non-zero
longitudinal flux-line energies attached to gluons (\ref{gg})(\ref{qg})
will tend to suppress their pair production, by whatever process, even
when
they are massless.

\subsection{Transverse.}

To a first approximation, the effect of $s$ and $t$ terms is to lift some of
the
helicity degeneracy. However they also mix sectors of different $n$
and lead to new divergences. The $t$ self-energy simply renormalises
$m_0$.
To illustrate the cancellation by the
$s$ self-energy we give below the integral
equations
for the $h=0$ sector restricting Fock space to at most one gluon, the
non-zero coefficients being $f_{+-},f_{-+},f_{+-+}$, and $f_{-+-}$ in
this case.
\begin{eqnarray}
M^2 f_{+-}(x_1,x_2) & = &
 m^2 \left( \frac{1}{x_1} + \frac{1}{x_2} \right) f_{+-}(x_1,x_2)
\nonumber \\
&& + {s^2 g^2 N_c\over \pi} \int_{0}^{\infty} {dy \over y}
\left\{ {1\over (x_1 -y)} + {1 \over (y+x_2)}\right\}  f_{+-}(x_1,x_2)
             \nonumber \\
& & - m s g \sqrt{ \frac{ N_c}{  \pi}} \int_{0}^{x_1}
        \frac{dy}{\sqrt{y}} \left(\frac{1}{x_1-y} - \frac{1}{x_1}
\right) f_{-+-}(x_1-y,y,x_2)  \nonumber    \\
& & - m sg\sqrt{ \frac{ N_c}{  \pi}} \int_{0}^{x_2}
        \frac{dy}{\sqrt{y}} \left(\frac{1}{x_2-y} - \frac{1}{x_2}
\right) f_{+-+}(x_1,y,x_2-y) + \dots \ ,\label{logdiv2}\\
& & \nonumber \\
 M^2 f_{-+-}(x_1,x_2,x_3) & = &
 m^2   \left( \frac{1}{x_1} + \frac{1}{x_3}
\right)f_{-+-}(x_1,x_2,x_3) \nonumber \\
& & - msg \sqrt{ \frac{ N_c}{  \pi}} \frac{1}{\sqrt{x_2}} \left(
        \frac{1}{x_1} - \frac{1}{x_2+x_1} \right) f_{+-}(x_1+x_2,x_3)
\nonumber
\\
& & +  m sg\sqrt{ \frac{ N_c}{  \pi}} \frac{1}{\sqrt{x_2}} \left(
        \frac{1}{x_3} - \frac{1}{x_2+x_3} \right) f_{-+}(x_1,x_2+x_3)
\nonumber \\
&& + {s^2 g^2 N_c \over \pi}
\int_{0}^{x_2 +x_3} dy
{f_{-+-}(x_1,y,x_3+x_2 -y)  \over (x_2 + x_3) \sqrt{x_2 y} } \nonumber
\\
&& +
{s^2 g^2 N_c \over \pi} \int_{0}^{x_1 +x_2} dy
{f_{-+-}(x_1 +x_2 -y ,y,x_3)  \over (x_2 + x_1) \sqrt{x_2 y} } + \dots
\label{nolog}
\end{eqnarray}
and the same equations with $\{+ \leftrightarrow -, s \leftrightarrow -s\}$.
Ellipses indicate purely longitudinal processes, dealt with in the
previous subsection.
The self-energy terms (curly brackets) together with the following two
integrals in (\ref{logdiv2}) appear divergent (they are). To see that
the divergences  cancel,  rewrite the $f_{-+-}$ integral singular at
anti-quark momentum $x_1-y = 0$ as
\begin{eqnarray}
 - m sg\sqrt{ \frac{ N_c}{  \pi}} & & \!\!\!\!  \left\{
  \int_{0}^{x_1} \frac{dy}{\sqrt{x_1}(x_1-y)}
    \left[
       f_{-+-}(x_1-y,y,x_2) -  f_{-+-}(0,x_1,x_2)
    \right] \right.  \nonumber  \\
 & & \left. -  \int_{0}^{x_1} \frac{dy}{x_1(\sqrt{x_1}+\sqrt{y})}
          f_{-+-}(x_1-y,y,x_2)  + \frac{f_{-+-}(0,x_1,x_2)}
                 {\sqrt{x_1}} \int_{0}^{x_1} \frac{dy}{x_1-y}
               \right\} \label{logdiv3}
\end{eqnarray}
to isolate a logarithmic divergence in the last integral of (\ref{logdiv3}).
 From the behaviour of (\ref{nolog}) as $x_1 \to 0$ one deduces the relation
\begin{equation}
f_{-+-}(0,x_2,x_3) =  \frac{sg}{m}
      \sqrt{ \frac{  N_c}{  \pi}} \frac{f_{+-}(x_2,x_3)}
                                      {\sqrt{x_2}},
 \label{relation1}
\end{equation}
to cancel the $\frac{1}{x_1}$ singularity (the longitudinal processes
don't contribute at this order), and therefore the divergent term in
(\ref{logdiv3}) is
\begin{equation}
  - \frac{s^2 g^2 N_c}{  \pi}
        \int_{0}^{x_1} \frac{dy}{x_1(x_1-y)}
             \hspace{1mm} f_{+-}(x_1,x_2),
\end{equation}
which is readily seen to cancel the anti-quark $s$ self-energy divergence.
The same argument goes through for the quark and other helicity.
Such cancelations may be understood also through 1-loop light-cone
perturbation
theory \cite{perry}.
In fact
the full integral equations imply the general endpoint conditions for
quarks:
\be
f_{\a \b  \dots \gamma \mp \pm}(x_1,\dots , x_{n-2},x_{n-1},0)
 =  \pm \frac{sg}{m} \sqrt{ \frac{ N_c}{x_{n-1}  \pi}}
f_{\a \b \dots \gamma \mp}(x_1,\dots ,
      x_{n-2},x_{n-1})
\eq
for all helicities, with a similar relation for anti-quarks.
These may be used to show in exactly the same way   the  cancelation
of  divergent integrals due to zero momentum (anti)quarks for every equation
i.e. involving an  arbitrary number of gluons,
since only the gluon neighbouring
the quark in the large $N_c$
colour contraction is involved.\footnote{Restricting to no more
than $n-2$ gluons, the $s$ self-energies are omitted for Fock states with $n-2$
gluons.} Since no other divergent integrals arise, the boundstate
equations are each finite.
We offer numerical evidence for this finiteness in the next section.
There is a possibility of divergence due to
there being an infinite number of equations however. In fact this is
to be expected for sufficiently large coupling constants from the
the well-known divergence of large-$N_c$ planar perturbation theory, which was
studied
in the light-front Hamiltonian formalism in refs.\cite{us,dk}.

\section{Numerical Solutions.}
The full boundstate equations may be solved numerically by discretizing the
momentum fractions as $x = m/K$ for integers $\{ m,K \}$ with $0 < m < K$, then
extrapolating to $K =\infty$ \cite{brod}. Here we will exhibit results
for one particular set of parameters: $m_B = 0$, $m^2 = 0.8$, $s=0.5$,
$t=0.15$, $g^2 N_c = \pi$;
the investigation of the full parameter space is left for
future
work.
A tractable problem is only obtained by further truncating the Fock
space
by hand.  We therefore extrapolated to $K=\infty$ the
discretised problem  with at most two transverse gluons, and checked
for finite $K$ that the error in restricting the number of gluons was
not large for the set of couplings used.
This restriction is the simplest such that all terms in
$P^-$
(\ref{minus}) are used.
Since the problem at finite
$K$ is
equivalent to finite matrix diagonalisation of the $M^2$ operator
\cite{brod},
 we employed a Lanczos algorithm whose implementation has been
described elsewhere \cite{us}. This algorithm is an iterative scheme
which
requires an input wavefunction close to an eigenfunction $\Psi$ if
convergence
is to  be good. Since the algorithm also preserves the
symmetries of the theory, a sensible input will consist of a couple of
valence quarks $\bar{q}q$
which
transform
in a definite way under symmetries of the discretised theory.

At finite $K$ the symmetry $C$ induced from charge conjugation
\be
C: \ a_{\pm ij} \rightarrow -a_{\pm ji} \ , \ b_{\pm i}
\leftrightarrow d_{\pm i}
\eq
remains exact, although parity $x^1 \to -x^1$
does not. In view of the difficulties
in measuring true parity, at this stage we content ourselves with a
classification according to the exact finite-$K$ symmetry of the momentum
fraction in the $\bar{q}q$ sector of Fock space
\be
P_1 \ : \ x \leftrightarrow 1-x \ .
\eq
In the continuum theory this $P_1$-parity is equivalent
to true parity in the limit of on-shell partons \cite{bhp}.
We take quarks and anti-quarks to have opposite
intrinsic $P_1$ by analogy with true parity.
We consider here the following valence combinations,
 classified by $|h|^{P_1 C}$:
\begin{eqnarray}
0^{-+} \ : &&\  d^{\d}_{+i} (x) b^{\d}_{-i} (1-x) - d^{\d}_{-i} (x)
b^{\d}_{+i} (1-x)|0> \\
0^{--} \ :&& \ d^{\d}_{+i} (x) b^{\d}_{-i} (1-x) +  d^{\d}_{-i} (x)
b^{\d}_{+i} (1-x)|0> \\
1^{--} \ :&& \ d^{\d}_{\pm i} (x) b^{\d}_{\pm i} (1-x)|0>  \ .
\end{eqnarray}
These couple strongly to the groundstate in each symmetry sector,
and are evidently analogues of the $J_z$ components of the
$\pi$ (or $\eta'$) and $\rho$ mesons,  if we identify $h\equiv J_z $. Hybrids
appear as excited $(*)$ states of them.

 From the eigenfunctions $\Psi$ one may compute unpolarized and
polarized structure functions
\begin{eqnarray}
Q(x)  =  <b^{\d}_{\a i} b_{\a i}(x) + d^{\d}_{\a i} d_{\a i} (x)> \ ,&
\Delta Q(x)  =  <\half\sum_{\a} {\rm sgn}(\a) [b^{\d}_{\a i} b_{\a i}(x) +
d^{\d}_{\a i} d_{\a i} (x)]> \\
G(x)  = <a^{\d}_{\a ij} a_{\a ij} (x)> \ , &
 \Delta G(x)  = <\sum_{\a}
{\rm sgn} \hspace{1mm}(\a) a^{\d}_{\a ij} a_{\a ij} (x) >
\end{eqnarray}
which trivially satisfy momentum and helicity sum rules
\begin{eqnarray}
\int_{0}^{1} dx \ x [Q(x) + G(x)] & =& 1 \\
\int_{0}^{1} dx \ [\Delta Q(x) + \Delta G(x)] & =& h \ .
\end{eqnarray}
It is also useful to define the integrated quantities $n_g =  \int
G(x) $, $ \Delta n_q  = \int \Delta Q(x)$, and the polarization
asymmetries $A_g(x) = \Delta G/G$, $A_q(x) = \Delta Q /Q$. Numerical
results for the chosen parameters
are displayed in Table 1 and Figures 1 and 2. There is clear
convergence of $M^2$, with $0^{-+}$ lightest (a robust feature).
The lightest `hybrid' occurs in the $1^{--}$
sector, suggesting that the hybrid $\r$ would be lightest.
The $Q$ and $G$ of $0^{--}$ and $1^{--}$ turn out to be
very similar to those of
$0^{-+}$ plotted. $G$ is small relative to $Q$ due to the rather small $s$
employed, neccesary to be sure of convergence. Note that even though
the
quarks are massive, $Q$ does not vanish at $x \to 0$, as a result of
eq.(\ref{relation1}). Both the
momentum and helicity of the $1^{--}$ are due to gluons at the level of
about $10\%$. The helicity asymmetries (fig. 2) also support this
helicity-momentum correlation, showing complete helicity alignment
of the parton as $x \to 1$, while totally disordered polarization
at small $x$.

\section{Conclusions.}

The $1+1$-dimensional reduced model of light-front large-$N_c$ QCD forms
an interesting extension of 't Hooft's original model for
two-dimensional mesons to include gluon degrees of
freedom and helicity. We have performed light-front quantization of
the normal modes and showed how divergences in
the infinite set of coupled boundstate equations are absorbed by
self-energies. A preliminary  analysis of
the meson boundstates was given and it is evidently desirable to make this
more comprehensive, varying the parameters and in particular checking
the true parity of states, which requires a more efficient
computer/code.
The way is now open to investigate non-perturbatively the
behaviour of scattering amplitudes involving the mesons and glueballs
constructed here and in ref.\cite{us} within the $1/N_c$ expansion.
It will be interesting to compare them with the hight-enrgy scattering
 processes of the
four-dimensional world.
\vspace{5mm}

\noindent {\bf Acknowledements:} We thank M.Burkardt, H-C.Pauli,
J.Paton,
and B. van de Sande for helpful discussions and  Prof. Pauli for his
hospitality at the MPI Heidelberg.

\vspace{10mm}
\begin{center}
FIGURE AND TABLE CAPTIONS
\end{center}
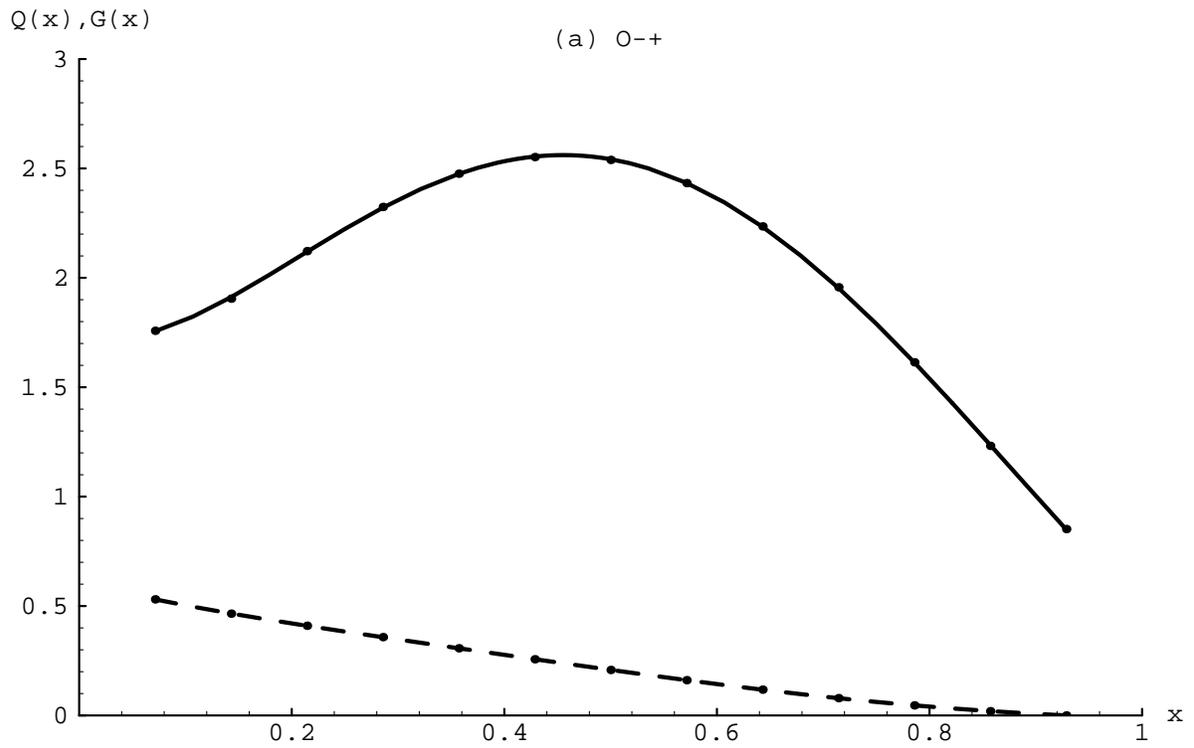
\noindent Figure 1. Quark and gluon momentum fraction distributions
for $0^{-+}$
and $1^{--}_{*}$; Solid lines $Q$; Broken lines $G$.

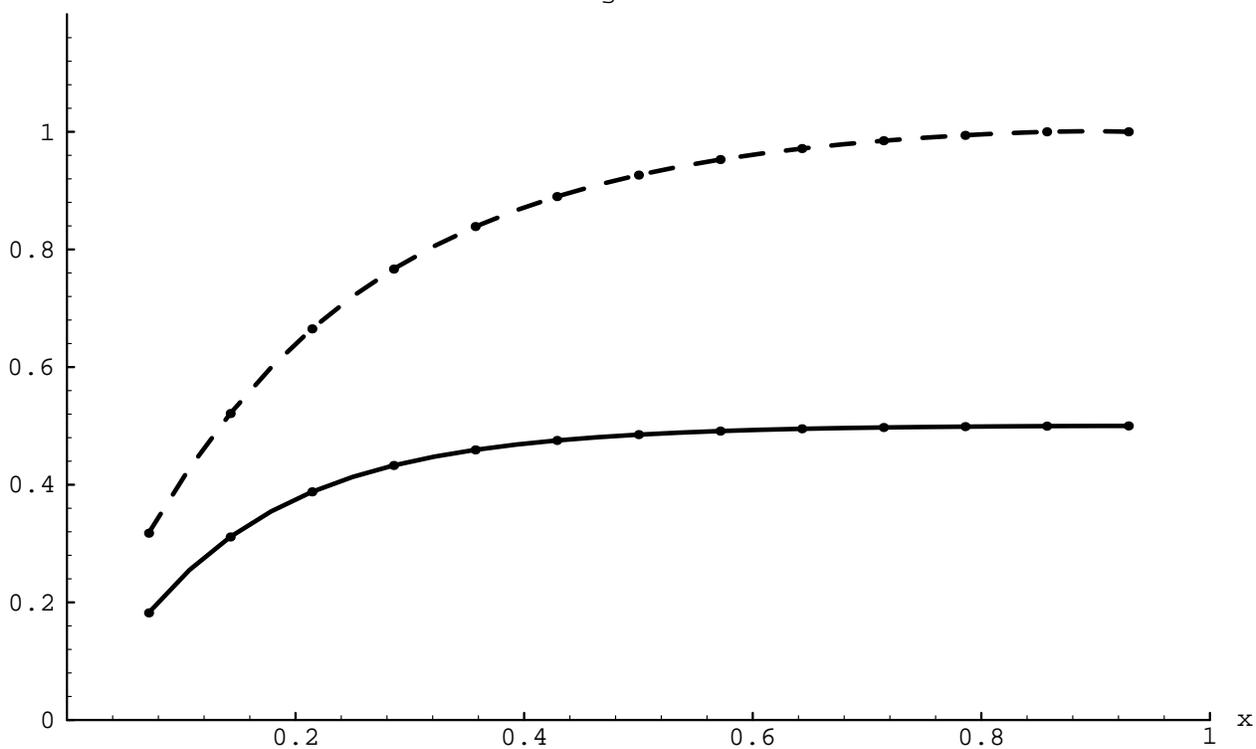
\noindent Figure 2. Quark and gluon helicity asymmetries for $1^{--}$;
solid line $A_q$; broken line $A_g$.

\noindent Table 1. Extrapolation of $M^2$ in cutoff $K$. $n_g$ and
$\Delta n_q$ are quoted at $K=14$.

\vspace{10mm}

\noindent Table.1
\medskip

\begin{tabular}{|c|c|c|c|c|c|} \hline
 $M^2 (K)$& $0^{-+}$ & $1^{--}$ & $0^{--}$ & $1^{--}_{*}$ \\ \hline
$M^2(12)$ & 5.3151 & 5.8393 & 6.2652 & 12.2958  \\ \hline
$M^2(13)$ & 5.3812 & 5.9243 & 6.3599 & 12.5327 \\ \hline
$M^2(14)$ & 5.4395 & 5.9992 & 6.4428 & 12.7455  \\ \hline
$M^2(15)$ & 5.4913 & 6.0655 & 6.5158 & 12.9412 \\ \hline
$M^2(16)$ & 5.5374 & 6.1247 & 6.5805 & 13.1229  \\ \hline
$M^2(17)$ & 5.5789 & -- & -- &   -- \\ \hline
$M^2(\infty)$ & 6.3 & 7.1 & 7.6 & 16.9 \\ \hline
$n_g$ & 0.23 & 0.17 & 0.08 & 0.81 \\ \hline
$\Delta n_q$ &0 &0.89 &0 &0.42  \\ \hline
\end{tabular}

\vfil

\begin{thebibliography}{9999}
\bibitem{hoof} G. 't Hooft, \NP {\bf B75} (1974) 461.
\bibitem{us} F. Antonuccio and S. Dalley,
preprint OUTP-9524P (hepth/9506456) \NP {\bf B} in press.
\bibitem{dk} S. Dalley and I.R. Klebanov, \PR {\bf D47} (1993) 2517;\\
K. Demeterfi, I.R. Klebanov, and G. Bhanot, \NP {\bf B418} (1994) 15.
\bibitem{su3} M.Burkardt and B. van de Sande, preprint MPI H-V38-1995
(hepth/9510104).
\bibitem{brod} H-C. Pauli and S.J. Brodsky, \PR {\bf D32} (1985) 1993
and 2001.
\bibitem{perry} A. Harindranath and R. Perry, \PR {\bf D43} (1991) 4051.
\bibitem{dk2} S. Dalley and I.R. Klebanov, \PL {\bf B298} (1993) 79.
\bibitem{zero} H-C.Pauli, A.C. Kalloniatis, and S.Pinsky, \PR {\bf
D52} (1995) 1176.
\bibitem{bhp} K. Hornbostel, Ph.D Thesis, SLAC report No. 333 (1988).
\end{thebibliography}
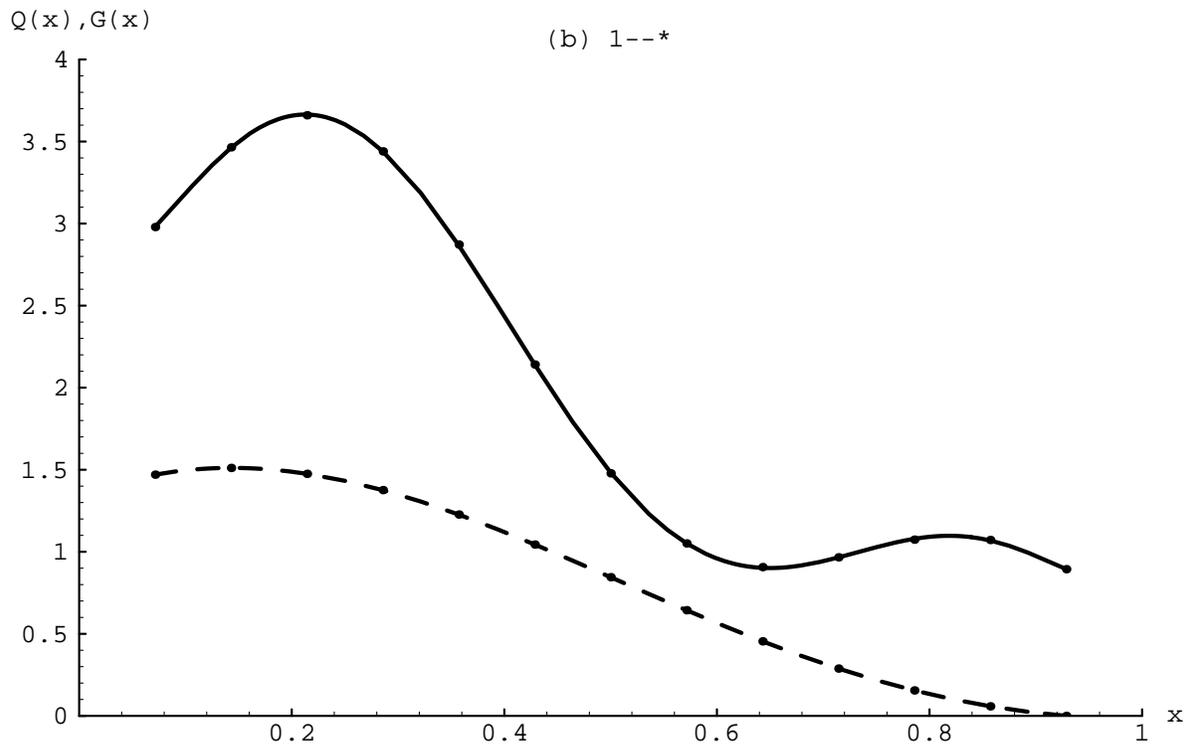
\end{document}